\documentclass[12pt]{article}
\usepackage{amsmath,amsfonts,amssymb}
\usepackage{cases}

\usepackage{indentfirst}
\makeatletter
\renewcommand{\@begintheorem}[2]{\begin{trivlist}
\item[\hspace{\labelsep}{\bfseries#1\ #2.}]\itshape}
\renewcommand{\@opargbegintheorem}[3]{\begin{trivlist}
\item[\hspace{\labelsep}{\bfseries#1\ #2\ (#3).}]\itshape}
\renewcommand{\@endtheorem}{\end{trivlist}}
\makeatother
\author{Sergey V.\,Smirnov\thanks{School of Mathematics and Statistics, University of Glasgow. E-mail: {\tt Sergey.Smirnov@glasgow.ac.uk}}}
\title{Darboux formulae for linear hyperbolic equations\\ in discrete case}

\def\pa{\partial}

\def\phi{\varphi}

\newcommand{\Ker}{\mathop{\rm Ker}\nolimits}

\newtheorem{proposition}{\sc Proposition}
\newtheorem{theorem}{\sc Theorem}
\newtheorem{lemma}{\sc Lemma}
\newtheorem{remark}{\sc Remark}
\newtheorem{example}{\sc Example}

\begin{document}

\maketitle
\begin{abstract}
In the second half of the 19th century Darboux obtained determinant formulae that provide the general solution for a linear hyperbolic second order PDE with finite Laplace series. These formulae played an important role in his study of the theory of surfaces and, in particular, in the theory of conjugate nets. During the last three decades discrete analogs of conjugate nets (Q-nets) were actively studied. Laplace series can be defined also for hyperbolic difference operators. We prove discrete analogs of Darboux formulae for discrete and semi-discrete hyperbolic operators with finite Laplace series.
\end{abstract}

\section{Introduction}

Linear hyperbolic second order differential operators had been extensively studied by Darboux~\cite{Da}, Goursat~\cite{Go} and their followers in the end of 19th century in relation to classical differential geometry: parameterisation of a hypersurface in 3-dimensional Euclidean space such that its second fundamental form is diagonal, is called a {\it conjugate net} in the theory of surfaces; it is known to satisfy a hyperbolic second order partial differential equation. Conjugate nets are considered in spaces of higher dimension both in affine and in projective settings~\cite{BS}.

Various problems in classical differential geometry can be reformulated in terms of solutions of linear hyperbolic equations. Therefore Darboux, Goursat and their followers were interested in developing techniques that allow to obtain exact solutions of such equations in explicit form. Some of them are based on the so-called {\it Laplace cascade method} that allows to express solutions of one hyperbolic equation in terms of solutions of another one whenever they are related by a finite sequence of {\it Laplace transformations}. If at least one of the {\it Laplace invariants} for a linear hyperbolic operator ${\mathcal L}$ is zero, then the general solution of the equation ${\mathcal L}\psi=0$ can be found explicitly. Combining this with the cascade method, one can obtain the general solution for hyperbolic equations that can be converted to the ones with a zero Laplace invariant by a finite number of Laplace transformations. There are two different Laplace transformations that can be applied to any hyperbolic operator with non-zero Laplace invariants. If application of a finite number of each of them leads to an operator with a zero Laplace invariant, then the initial operator is said to have {\it finite Laplace series}. There are several ways to express the general solution of a hyperbolic PDE with finite Laplace series. Darboux~\cite{Da} gives an elegant way to represent the general solution of all hyperbolic equations having finite Laplace series in terms of determinants that contain an appropriate number of arbitrary functions of one variable ({\it the Darboux determinant formula}). From a geometrical point of view, Laplace transformation is a transformation of a surface with a conjugate coordinate net. Thus the Darboux formula gives a nice representation of all conjugate nets with finite Laplace series in a convenient form. The Darboux formulae, the Laplace cascade method, and related things in the theory of integrable systems are discussed from a modern perspective in~\cite{GT}, but unfortunately this book is available only in Russian.

During the last three decades classical differential geometry has been discretised by Bobenko, Suris, Doliwa, Schief and many others. Exactly as in the continuous case, linear hyperbolic second order difference operator defines a {\it Q-net}, which is a discrete analog of a conjugate net~\cite{BS}. Laplace invariants are also defined in the discrete case~\cite{DN,AS,Ni1}. Discrete Laplace transformations provide transformations of $Q$-nets~\cite{Do1}. Semi-discrete analogs of conjugate nets, where one of the independent variables is continuous and the other one is discrete, had also been considered by specialists in discrete differential geometry~\cite{Mu}. Although the discrete case is rather similar to the continuous one, discrete analogs of the Darboux formulae seem to be missing in the existing literature. This paper is aimed at filling this gap.

Classical Laplace transformations play an important role not only in the theory of surfaces: they are closely related to famous integrable systems such as the two-dimensional Toda lattice and the Sine-Gordon equation. Laplace invariants of hyperbolic operators in a Laplace series satisfy nonlinear hyperbolic equations that are equivalent to the Toda system; finite Laplace series corresponds to the so-called {\it open Toda lattice} (also known as {\it $A$-series Toda lattice}) and the Sine-Gordon equation corresponds to the simplest periodic closure of Laplace series. Semi-discrete and entirely discrete versions of the two-dimensional Toda lattice are related to (semi)-discrete Laplace transformations in the same way as in the continuous case~\cite{AS} and are known to be Darboux integrable~\cite{Sm1}. 

In the continuous case symmetry reductions of the open Toda lattice lead to exponential systems corresponding to the Cartan matrices of $B$- and $C$-type~\cite{Le,ShYa}. From the point of view of the Laplace transformations, these reductions correspond to the Laplace series of the Goursat equation and the Moutard equation respectively~\cite{Da,GT}. From a geometrical point of view, B-type reduction corresponds to a special type of conjugate nets known as 
{\it K\"onigs nets}. Darboux also provides a procedure to describe the general solutions of these equations in the case when their Laplace series is finite~\cite{Da}. In the discrete case the situation with symmetry reduction is more complicated: $C$-series Toda lattice is known to be a reduction on the $A$-series lattice~\cite{Ha,Sm2}, but for $B$-series systems this is not clear. Moreover, it is not known whether (semi)-discrete $B$-series Toda systems introduced in~\cite{HZY,GHY} within the frame of a general approach to discretisation of exponential systems are related to the discrete version of the Moutard equation that is studied in~\cite{NS}--\cite{DGNS} and to discrete K\"onigs nets~\cite{BS}. We leave these questions as well as obtaining the discrete analog of the Darboux procedure for finding general solutions of the reduced systems in explicit form for future research.

In Section~\ref{lapl}, we review the basic notions such as the Lalplace invariants and the Laplace transformations. In Section~\ref{sectdarb}, we discuss the Darboux formulae for the general solution of a hyperbolic equation with a finite Laplace series in the continuous case. In Section~\ref{sectdlapl}, we recall the notions of the Laplace invariants and the Laplace transformations in the discrete case. Section~\ref{discr} contains the discrete analogs of the Darboux formulae and their detailed proof. In Section~\ref{semidiscr}, we provide semi-discrete versions of these formulae.

\section{Laplace invariants and Laplace series}\label{lapl}

In this Section, we review the main notions and facts concerning the Laplace invariants, the Laplace transformations and the Laplace series in the continuous case that we need to prove discrete analogs of the Darboux formulae. All propositions, theorems and statements in this Section are either trivial or can be found in~\cite{Da} with detailed proofs.

Consider linear hyperbolic operator
\begin{equation}
\label{bbb}
{\mathcal L}=\pa_x\pa_y+a\pa_x+b\pa_y+c.
\end{equation}
The functions $k=c-ab-a_x$ and $h=c-ab-b_y$ are called {\it the Laplace invariants} of~(\ref{bbb}). They are invariant under gauge 
transformations ${\mathcal L}\mapsto\bar{\mathcal L}=\omega^{-1}{\mathcal L}\omega$, where $\omega=\omega(x,y)$ is an arbitrary
function, and control factorizability of such an operator: since
$$
{\mathcal L}=(\pa_x+b)(\pa_y+a)+k=(\pa_y+a)(\pa_x+b)+h,
$$
the operator ${\mathcal L}$ is factorisable if and only if at least one of its Laplace invariants is zero. Straightforward calculation shows that if $h=0$, then the general solution of ${\mathcal L}\psi=0$ has the form
\begin{equation}
\label{ddd}
\psi(x,y)=\left(\int\left(X(x) e^{\int b(x,y)dx-\int a(x,y)dy}\right)dx+Y(y)\right)e^{-\int b(x,y)dx}
\end{equation}
and if $k=0$, then the general solution has the form
\begin{equation}
\label{eee}
\psi(x,y)=\left(\int\left(Y(y) e^{\int a(x,y)dy-\int b(x,y)dx}\right)dy+X(x)\right)e^{-\int a(x,y)dy},
\end{equation}
where $X$ and $Y$ are arbitrary functions of one variable.

Two linear hyperbolic differential operators
$$
{\mathcal L}=\pa_x\pa_y+a\pa_x+b\pa_y+c\quad\hbox{and}\quad\hat{\mathcal L}=\pa_x\pa_y+\hat a\pa_x+\hat b\pa_y+\hat c
$$
are related by a {\it Darboux--Laplace transformation}, if there exist first order operators ${\mathcal D}=\alpha\pa_x+\beta\pa_y+\gamma$ and
$\hat{\mathcal D}=\hat\alpha\pa_x+\hat\beta\pa_y+\hat\gamma$ such that the following relation is satisfied:
\begin{equation}
\label{dlt}
\hat{\mathcal L}\circ {\mathcal D}=\hat{\mathcal D}\circ {\mathcal L}.
\end{equation}
Two particular cases ${\mathcal D}=\pa_x+b$ and ${\mathcal D}=\pa_y+a$ correspond to classical {\it Laplace transformations} that were widely used by Darboux and Goursat in their research on the theory of surfaces. In both cases the coefficients of operators $\hat{\mathcal L}$ and $\hat{\mathcal D}$ are uniquely defined in terms of the coefficients of ${\mathcal L}$. The main property of Darboux-Laplace transformations is that the transformation operator 
${\mathcal D}$ maps the kernel of the operator ${\mathcal L}$ into the kernel of $\hat{\mathcal L}$: if $\psi\in\Ker{\mathcal L}$, then relation~(\ref{dlt}) implies that ${\mathcal D}\psi\in\Ker\hat{\mathcal L}$. There exists a wide class of Darboux--Laplace transformations for linear hyperbolic operators different from the classical Laplace transformations (see~\cite{She1,She2}), but we will consider only classical transformations throughout this paper. The following proposition is verified straightforwardly.
\begin{proposition}
Any Laplace transformation ${\mathcal L}\mapsto\hat{\mathcal L}$ defined by the transformation ope\-ra\-tor ${\mathcal D}=\pa_x+b$ is invertible if $h\ne 0$. The inverse transformation $\hat{\mathcal L}\mapsto{\mathcal L}$ is given by the operator $-h^{-1}(\pa_y+a)$. Similarly,
the transformation defined by the operator $\pa_y+a$ is invertible if $k\ne 0$, and the inverse transformation is given by $-k^{-1}(\pa_x+b)$.
\end{proposition}

Consider a hyperbolic operator
$$
{\mathcal L}_0=\pa_x\pa_y+a_0\pa_x+b_0\pa_y+c_0
$$
and apply the Laplace transformation ${\mathcal D_0}=\pa_x+b_0$:
$$
{\mathcal L}_1{\mathcal D}_0={\mathcal D}_1{\mathcal L}_0\quad\hbox{where}\quad {\mathcal L}_1=\pa_x\pa_y+a_1\pa_x+b_1\pa_y+c_1.
$$
Apply the Laplace transformation ${\mathcal D_1}=\pa_x+b_1$ to the operator ${\mathcal L}_1$, etc. This procedure produces the sequence of hyperbolic operators starting from ${\mathcal L}_0$ and such that any two neighbouring operators are related by a Laplace transformation.
The second Laplace transformation ${\mathcal D}'_0=\pa_y+a_0$ allows the continue the sequence to the left:
$$
{\mathcal L}_{-1}{\mathcal D}'_0={\mathcal D}'_1{\mathcal L}_0,
$$
etc. This leads to the sequence
\begin{equation}
\label{mmm}
\dots\stackrel{{\mathcal D}'_j}{\longleftarrow}{\mathcal L}_{-j}\stackrel{{\mathcal D}'_{j-1}}{\longleftarrow}{\mathcal L}_{-j+1}\stackrel{{\mathcal D}'_{j-2}}
{\longleftarrow}\dots\stackrel{{\mathcal D}'_1}{\longleftarrow}{\mathcal L}_{-1}\stackrel{{\mathcal D}'_0}{\longleftarrow}{\mathcal L}_0\stackrel{{\mathcal D}_0}
{\longrightarrow}{\mathcal L}_1\stackrel{{\mathcal D}_1}{\longrightarrow}\dots\stackrel{{\mathcal D}_{j-1}}{\longrightarrow}{\mathcal L}_j\stackrel{{\mathcal D}_j}
{\longrightarrow}{\mathcal L}_{j+1}\stackrel{{\mathcal D}_{j+1}}{\longrightarrow}\dots,
\end{equation}
which is called {\it the Laplace series} of the operator ${\mathcal L}_0$. If some operator ${\mathcal L}_r$ in its positive part has zero $h$-invariant, then the sequence terminates at ${\mathcal L}_r$ since this operator is factorisable. Similarly, if ${\mathcal L}_{-s}$ has zero $k$-invariant for some $s\geqslant 0$, then sequence~(\ref{mmm}) terminates from the left. The Laplace series is said to be finite if it terminates from both sides.

Laplace invariants of any three consecutive operators in the Laplace series satisfy the relations
$$
k_{j+1}=h_j,\quad h_{j+1}=2h_j-h_{j-1}+(\ln h_j)_{xy}.
$$
The second of these relations is one of the forms of the {\it two-dimensional Toda lattice}.

The theorem below generalises formulae~(\ref{ddd}),~(\ref{eee}). It follows from the fact that each operator in the Laplace series is obtained from the previous one by a Laplace transformation, which maps the kernel of the initial operator into the kernel of the next one.
\begin{theorem}
If the Laplace series of ${\mathcal L}_0$ terminates in the term ${\mathcal L}_r$ for some $r\in\mathbb N$, then the general solution of the equation 
${\mathcal L}_0\psi=0$ is given by
\begin{equation}
\label{nnn}
\psi_0(x,y)=-\frac{1}{h_0}(\pa_y+a_0)\left(-\frac{1}{h_1}(\pa_y+a_1)\left(\dots\left(-\frac{1}{h_{r-1}}(\pa_y+a_{r-1})\right)\dots\right)\right)\psi_r(x,y),
\end{equation}
where $\psi_r$ has the form
\begin{equation}
\label{ooo}
\psi_r(x,y)=\left(\int\left(X(x) e^{\int b_r(x,y)dx-\int a_r(x,y)dy}\right)dx+Y(y)\right)e^{-\int b_r(x,y)dx}
\end{equation}
and  $X$ and $Y$ are arbitrary functions of one variable. If the Laplace series of ${\mathcal L}_0$ terminates in the term ${\mathcal L}_{-s}$ for some $s\in\mathbb N$, then the general solution is given by
\begin{equation}
\label{ppp}
\psi_0(x,y)=-\frac{1}{k_0}(\pa_x+b_0)\left(-\frac{1}{k_{-1}}(\pa_x+b_{-1})\left(\dots\left(-\frac{1}{k_{-s+1}}(\pa_x+b_{-s+1})\right)\dots\right)\right)\psi_{-s}(x,y),
\end{equation}
where $\psi_{-s}$ is defined as follows:
$$
\psi_{-s}(x,y)=\left(\int\left(Y(y) e^{\int a_{-s}(x,y)dy-\int b_{-s}(x,y)dx}\right)dy+X(x)\right)e^{-\int a_{-s}(x,y)dy}.
$$
\end{theorem}
\begin{remark}
\rm
The possibility to obtain the general solution in form~(\ref{nnn}) or~(\ref{ppp}) is a consequence of the fact that an operator with zero Laplace invariant can be factorised. The method of consecutive application of Laplace transformations to a hyperbolic operator with the Laplace series that terminates at least in one direction is called the {\it Laplace cascade method}.
\end{remark}

\section{Darboux formulae for general solution}\label{sectdarb}

formulae~(\ref{nnn}) and~(\ref{ppp}) from the previous Section allow one to obtain the general solution of a particular hyperbolic equation in explicit form given that its Laplace series is finite at least in one direction. But on the one hand they include the integration of an arbitrary function, and on the other hand they do not provide a suitable description of all hyperbolic equations with finite Laplace series, which would be interesting from a geometrical point of view. In this Section, we review the Darboux formulae from~\cite{Da} that give the general solution in a more suitable form.

\begin{theorem}
If the Laplace series of ${\mathcal L}_0$ is finite, $h_r=k_{-s}=0$ for some $r,s\geqslant 0$, then the general solution of the equation 
${\mathcal L}_0\psi=0$  is given by
\begin{equation}
\label{uuu}
\psi(x,y)=A_0 X+A_1 X'+\dots+A_s X^{(s)}+B_0 Y+B_1 Y'+\dots+B_r Y^{(r)},
\end{equation}
where $X(x)$ and $Y(y)$ are arbitrary functions of one variable, $A_j(x,y)$ and $B_j(x,y)$ are some particular functions depending
the coefficients of the initial equation (primes and superscripts denote derivatives).

If a hyperbolic equation ${\mathcal L}_0\psi=0$ admits a solution of the form~(\ref{uuu}), then its Laplace series is finite: $h_r=k_{-s}=0$.
\end{theorem}
\begin{remark}
\rm
Not every expression of form~(\ref{uuu}) defines the general solution for some hyperbolic equation with finite Laplace series since there are restrictions on the coefficients $A_i$, $B_j$. Indeed, consider the simplest case of an operator with zero Laplace invariants $h_0=k_0=0$. Hence, formula~(\ref{uuu}) takes the form 
$\psi=A X+B Y$. Differentiate it with respect to $x$ and $y$ and substitute into the initial hyperbolic equation. Since $X$ and $Y$ are arbitrary, this yields
\begin{equation}
\label{fff}
a_0=-(\ln A)_y,\quad b_0=(\ln B)_x,\quad (\ln A)_{xy}=(\ln B)_{xy}.
\end{equation}
The third relation shows that functions $A$ and $B$ could not be chosen at random.
\end{remark}

Complete description of all restrictions that should be imposed on the coefficients of~(\ref{uuu}) is given by the {\it Darboux determinant formula}.
\begin{theorem}
Let ${\mathcal L}_0$ be a hyperbolic operator with finite Laplace series, where $h_r=k_{-s}=0$ for some $r,s\geqslant 0$. Then there exists a function $\omega$ such that the general solution of ${\mathcal L}_0\psi=0$ is given by
\begin{eqnarray}
\label{aaa}
\psi(x,y)=\omega(x,y)\cdot\left|
\begin{array}{cccccccc}
X & X' & \dots & X^{(s)} & Y & Y' & \dots & Y^{(r)}\\
\xi_1 & \xi'_1 & \dots & \xi^{(s)}_1 & \eta_1 & \eta'_1 & \dots & \eta^{(r)}_1\\
\vdots &&& \vdots & \vdots &&& \vdots\\
\xi_{r+s+1} & \xi'_{r+s+1} & \dots & \xi^{(s)}_{r+s+1} & \eta_{r+s+1} & \eta'_{r+s+1} & \dots & \eta^{(r)}_{r+s+1}\\
\end{array}
\right|
\end{eqnarray}
where $X(x)$ and $Y(y)$ are arbitrary functions of one variable and 
\begin{equation}
\label{arb}
\xi_1 (x),\xi_2 (x),\dots,\xi_{r+s+1}(x),\eta_1(y),\eta_2 (y),\dots,\eta_{r+s+1}(y)
\end{equation}
are some specific functions of one variable depending on the coefficients of the operator ${\mathcal L}_0$.\\

For any two families of linearly independent functions of one variable~(\ref{arb}) and for any function $\omega$ there exists a hyperbolic differential operator 
${\mathcal L}_0$ with finite Laplace series that terminates at the terms $r$ and $-s$, such that formula~(\ref{aaa}) provides the general solution of 
${\mathcal L}_0\psi=0$.
\end{theorem}
\begin{example}
\rm
In the simplest case $h_0=k_0=0$ formula~(\ref{aaa}) gives the general solution 
$$
\psi(x,y)=\eta(y)X(x)-\xi(x)Y(y).
$$ 
Note that in this case 
$a_0=-(\ln\eta)_y$, $b_0=-(\ln\xi)_x$, and the compatibility condition from~(\ref{fff}) is satisfied since $\xi$ and $\eta$ are functions of one variable.
\end{example}
\begin{remark}
\rm
From a geometrical point of view determinant formula~(\ref{aaa}) provides a large stock of surfaces with conjugate coordinate nets and with finite Laplace series.
\end{remark}
\begin{remark}\label{remcoef}
\rm
It is also possible to find effectively the coefficients of hyperbolic operator ${\mathcal L}_0$ that is constructed given families~(\ref{arb}) of arbitrary functions. Indeed, the expansion of determinant~(\ref{aaa}) along the first row allows to express the coefficients $A_0,\dots,A_r,B_0,\dots,B_s$ in~(\ref{uuu}) in terms of functions~(\ref{arb}). The coefficients $a_0$, $b_0$ and $c_0$, in turn, can be expressed in terms of $A_i$, $B_j$ in an explicit form: these formulae can be found in~\cite{Da}. We will only give their analog in the discrete case.
\end{remark}

\section{Laplace invariants in discrete case}\label{sectdlapl}

In this Section, we aggregate basic facts on the Laplace invariants, Laplace transformations, and the cascade method both in the purely discrete and in the semi-discrete cases. Although to the best of our knowledge some of the propositions below are not contained in the existing literature in explicit form, their proofs are either trivial or similar to the proofs of analogous propositions in the continuous case. Therefore we give only statements without proofs in this Section.

In the purely discrete case consider a sequence of hyperbolic difference operators
$$
{\cal L}_j=T_n T_m+a_j T_n+b_j T_m+c_j,
$$
where functions $a_j$, $b_j$ and $c_j$ are defined on the $(n,m)$-lattice and $T_n$, $T_m$ are the shift operators:
$$
T_n\psi_{n,m}=\psi_{n+1,m}\quad T_m\psi_{n,m}=\psi_{n,m+1}.
$$
Clearly, the operator ${\cal L}_j$ can be factorised in two different ways:
\begin{multline}
\nonumber
{\cal L}_j=(T_n+b_{j,n,m})(T_m+a_{j,n-1,m})+a_{j,n-1,m}b_{j,n,m}k_{j,n,m}=\\
=(T_m+a_{j,n,m})(T_n+b_{j,n,m-1})+a_{j,n,m}b_{j,n,m-1}h_{j,n,m},
\end{multline}
where
$$
k_{j,n,m}=\frac{c_{j,n,m}}{a_{j,n-1,m}b_{j,n,m}}-1\quad\hbox{and}\quad h_{j,n,m}=\frac{c_{j,n,m}}{a_{j,n,m}b_{j,n,m-1}}-1
$$
are the {\it Laplace invariants} of difference operators ${\cal L}_j$. Suppose any two neighbouring operators ${\cal L}_j$ and ${\cal L}_{j+1}$ are related
by the {\it Laplace transformation} defined by ${\cal D}_j=T_n+b_{j,n,m-1}$. Then the Laplace invariants 
satisfy the {\it purely discrete two-dimensional Toda lattice}~\cite{AS}:
$$
k_{j+1,n,m}=h_{j,n,m},\quad\frac{h_{j,n+1,m}h_{j,n,m-1}}{h_{j,n,m}h_{j,n+1,m-1}}=\frac{(1+h_{j+1,n,m})(1+h_{j-1,n+1,m-1})}{(1+h_{j,n,m})(1+h_{j,n+1,m-1})}.
$$
Darboux--Laplace transformations of a more general form in the discrete case were considered in~\cite{Sm3}.
\begin{proposition}
Two hyperbolic second order difference operators ${\mathcal L}$ and $\bar{\mathcal L}$ have the same Laplace invariants $\bar k=k$ and $\bar h=h$ if and only if they are related by a gauge transformation of the form $\bar{\mathcal L}=\omega^{-1}_{n+1,m+1}{\mathcal L}\omega_{n,m}$ for some function $\omega_{n,m}$.
\end{proposition}
\begin{remark}
\rm
Note that in the discrete case the Laplace invariants are defined not quite si\-mi\-lar\-ly to the continuous case: there is division by $a_{j,n,m}$ and by $b_{j,n,m}$. This comes from the fact that functions
$$
\tilde k_{j,n,m}=c_{j,n,m}-a_{j,n-1,m}b_{j,n,m},\quad\tilde h_{j,n,m}=c_{j,n,m}-a_{j,n,m}b_{j,n,m-1}
$$
are not invariant with respect to gauge transformations. 
\end{remark}
\begin{remark}
\rm
Throughout the whole paper we will assume that coefficients $a_{0,n,m}$ and $b_{0,n,m}$ of the initial operator are non-zero on the whole lattice. As a consequence, coefficients of all operators in the series also appear to be non-zero (this follows from the relations on the coefficients of neighbouring operators that are deduced from~(\ref{dlt})). Hence the Laplace invariants are well-defined. The assumption we made is rather natural from a geometrical point of view: hyperbolic equations that we consider define $Q$-nets, that is, discrete surfaces made of quadrilaterals (see~\cite{BS}), and if a coefficient of such hyperbolic operator turns into zero, then the corresponding quadrilateral degenerates into a triangle.
\end{remark}

In the continuous case, we can use the cascade method for finding the general solution of a hyperbolic equation whenever its Laplace series terminates. But in order to do this we need to find the general solution of the corresponding hyperbolic equation with zero Laplace invariant. In the discrete case, there is a problem that the apparatus of difference equations is less developed than the apparatus of differential equations: it is possible to express the general solution of a factorisable hyperbolic differential equation in compact form~(\ref{ddd}) or~(\ref{eee}) using integration, although appropriate notation for similar situation in the discrete case does not exist in mathematics. Indeed, using the representation
$$
{\mathcal L}=(T_n+b_{n,m})(T_m+a_{n-1,m})
$$
of a difference operator for which $k_{n,m}=0$, one can denote $\phi_{n,m}=\psi_{n,m+1}+a_{n-1,m}\psi_{n,m}$ and consider non-homogeneous difference equation 
$\phi_{n+1,m}+b_{n,m}\phi_{n,m}=0$ for the new unknown function $\phi_{n,m}$. For any integers $n,m$ the value of $\phi_{n,m}$ can be recursively expressed in terms of the coefficients and the initial data $\phi_{0,m}$. Consecutively, given $\phi_{n,m}$, one can solve the first order difference equation for 
$\psi_{n,m+1}$ and write down the general solution in terms of the initial data $\psi_{k,0}$ and $\psi_{0,k}$. For example, for positive values of $m$ this expression has the form
\begin{equation}
\label{xxx}
\psi_{n,m}=\alpha_{n,m}\psi_{n,0}+\gamma^0_{n,m}\psi_{0,0}+\gamma^1_{n,m}\psi_{0,1}+\gamma^2_{n,m}\psi_{0,2}+\dots+\gamma^m_{n,m}\psi_{0,m},
\end{equation}
where $\alpha$, $\gamma^k$ are some coefficients. Unfortunately, the particular form of this expression is rather ugly and depends on the signs of integers $n$ and $m$. Therefore we do not make an attempt to give this formula explicitly and only emphasise that there exists an effective way to express the general solution in terms of the initial data $N_n=\psi_{n,0}$ and $M_m=\psi_{0,m}$ once the corresponding difference operator is factorisable. Note that expression~(\ref{xxx}) is local in arbitrary function $N_n$ and non-local in arbitrary function $M_m$ (i.e. it depends on a variable number of its shifts). Similarly, if $h_{n,m}=0$, then the general solution can be expressed in the form
\begin{equation}
\label{zzz}
\psi_{n,m}=\beta_{n,m}\psi_{0,m}+\delta^0_{n,m}\psi_{0,0}+\delta^1_{n,m}\psi_{1,0}+\delta^2_{n,m}\psi_{2,0}+\dots+\delta^n_{n,m}\psi_{n,0}.
\end{equation}

Another way to represent the general solution of a hyperbolic difference equation with factorisable operator is given by
\begin{proposition}\label{propsol}
If ${\mathcal L}=(T_n+b_{n,m})(T_m+a_{n-1,m})$, then the general solution of ${\mathcal L}\psi=0$ has the form $\theta_{n,m}+\theta^0_{n,m}$, 
where $\theta_{n,m}$ is the general solution of homogeneous equation $(T_m+a_{n-1,m})\psi_{n,m}=0$, $\theta^0_{n,m}$ is a particular solution of non-homogeneous equation $(T_m+a_{n-1,m})\psi_{n,m}=\phi_{n,m}$ and $\phi$ ranges the space $\Ker(T_n+b_{n,m})$.

If ${\mathcal L}=(T_m+a_{n,m})(T_n+b_{n,m-1})$, then the general solution of ${\mathcal L}\psi=0$ has the form $\chi_{n,m}+\chi^0_{n,m}$, 
where $\chi_{n,m}$ is the general solution of homogeneous equation $(T_n+b_{n,m-1})\psi_{n,m}=0$, $\chi^0_{n,m}$ is a particular solution of non-homogeneous equation $(T_n+b_{n,m-1})\psi_{n,m}=\tau_{n,m}$ and $\tau$ ranges the space $\Ker(T_m+a_{n,m})$.
\end{proposition}
\begin{proposition}
The Laplace transformation ${\mathcal L}\mapsto\hat{\mathcal L}$ defined by the ope\-ra\-tor ${\mathcal D}=T_n+b_{n,m-1}$ is invertible if
$a_{n,m} b_{n,m-1} h_{n,m}\ne 0$. The inverse transformation $\hat{\mathcal L}\mapsto{\mathcal L}$ is given by the transformation operator
$-(a_{n,m}b_{n,m-1} h_{n,m})^{-1}(T_m+a_{n,m})$. Similarly, the transformation defined by the operator $T_m+a_{n-1,m}$ is invertible if $a_{n-1,m} b_{n,m}k_{n,m}\ne 0$,
and the inverse transformation is given by $-(a_{n-1,m} b_{n,m}k_{n,m})^{-1}(T_n+b_{n,m})$.
\end{proposition}
\begin{remark}\label{reminv}
\rm
The Laplace transformations $T_n+b_{n,m-1}$ and $T_m+a_{n,m}$ are not inverse to each other on the level of particular functions: we need to multiply by a coefficient in order to obtain the inverse transformation. But if everything is considered up to multiplication by a function $\omega_{n,m}$, then this difference is not essential. Therefore in such a situation we will not specify the particular form of this coefficient and will think of these transformations as being inverse to each other.
\end{remark}
\begin{theorem}
If the Laplace series of ${\mathcal L}_0$ terminates in the term ${\mathcal L}_r$ for some $r\in\mathbb N$, then the general solution of the equation 
${\mathcal L}_0\psi=0$ is given by
\begin{multline}
\label{ttt}
\psi_{0,n,m}=-\frac{1}{a_{0,n,m}b_{0,n-1,m}h_{0,n,m}}(T_m+a_{0,n,m})\left(-\frac{1}{a_{1,n,m}b_{1,n-1,m}h_{1,n,m}}(T_m+a_{1,n,m})\dots\right.\\
\dots\left.\left(-\frac{1}{a_{r-1,n,m}b_{r-1,n-1,m}h_{r-1,n,m}}(T_m+a_{r-1,n,m})\right)\dots\right)\psi_{r,n,m},
\end{multline}
where $\psi_r$ is the general solution of a factorisable equation ${\mathcal L}_r\psi=0$.

If the Laplace series of ${\mathcal L}_0$ terminates in the term ${\mathcal L}_{-s}$ for some $s\in\mathbb N$, then the general solution is given by
\begin{multline*}
\psi_{0,n,m}=-\frac{1}{a_{0,n,m-1}b_{0,n,m}h_{0,n,m}}(T_n+b_{0,n,m})\left(-\frac{1}{a_{-1,n,m-1}b_{-1,n,m}h_{-1,n,m}}(T_n+b_{1,n,m})\dots\right.\\
\dots\left.\left(-\frac{1}{a_{-s+1,n,m-1}b_{-s+1,n,m}h_{-s+1,n,m}}(T_n+b_{-s+1,n,m})\right)\dots\right)\psi_{-s,n,m},
\end{multline*}
where $\psi_{-s}$ is the general solution of a factorisable equation ${\mathcal L}_{-s}\psi=0$.
\end{theorem}

Consider a sequence of hyperbolic differential-difference operators
\begin{equation}
\label{aff}
{\cal L}_j=\pa_x T_n+a_{j,n}\pa_x+b_{j,n} T_n+c_{j,n},
\end{equation}
where $a_{j,n}$, $b_{j,n}$ and $c_{j,n}$ are functions depending on discrete variable $n\in\mathbb Z$ and on continuous variable $x\in\mathbb R$
and where $T$ is the shift operator: $T_n\psi_n (x)=\psi_{n+1}(x)$. The operator ${\cal L}_j$ can be rewritten in two different ways,
$$
{\cal L}_j=(\pa_x+b_{j,n})(T+a_{j,n})+a_{j,n}k_{j,n}=(T+a_{j,n})(\pa_x+b_{j,n-1})+a_{j,n}h_{j,n},
$$
where
$$
k_{j,n}=\frac{c_{j,n}}{a_{j,n}}-(\ln a_{j,n})'_x-b_{j,n}\quad\hbox{and}\quad h_{j,n}=\frac{c_{j,n}}{a_{j,n}}-b_{j,n-1}
$$
are the {\it Laplace invariants} of differential-difference operator ${\cal L}_j$. Similarly to the
continuous case operator~(\ref{aff}) can be factorised if and only if at least one of its Laplace invariants vanishes. 

Suppose any two neighbouring operators ${\cal L}_j$ and ${\cal L}_{j+1}$ are related by a {\it Laplace transformation}, that is, they satisfy relation~(\ref{dlt}),
where ${\cal D}_j=\pa_x+b_{j,n-1}$. This operator relation can be rewritten in terms of coefficients:
\begin{eqnarray}
\nonumber
\left\lbrace
\begin{array}{l}
k_{j+1,n}=h_{j,n}\\
\left(\ln\frac{h_{j,n}}{h_{j,n+1}}\right)'_x=h_{j+1,n+1}-h_{j,n+1}-h_{j,n}+h_{j-1,n}
\end{array}
\right..
\end{eqnarray}
These equations are one of the forms of the semi-discrete two-dimensional Toda lattice~\cite{AS}. Laplace transformations in the variable $n$ are defined by difference operators ${\mathcal D}'_j=T_n+a_{j,n}$.
\begin{remark}
\rm
In the semi-discrete case the division by $a_{j,n}$ in the definition of the Laplace invariants in also necessary because the functions
$$
\tilde k_{j,n}=c_{j,n}-a_{j,n,x}-a_{j,n}b_{j,n},\quad\tilde h_{j,n}=c_{j,n}-a_{j,n}b_{j,n-1}
$$
are not invariant with respect to gauge transformations of the form ${\mathcal L}\to\omega^{-1}_{n+1}{\mathcal L}\omega_n$. Therefore in this case we also assume that the coefficients $a_{0,n}$ are non-zero. Since operator relation~(\ref{dlt}) yields the condition $a_{j,n}=a_{j+1,n}$, the coefficients of all operators in the lattice do not turn into zero, and hence the Laplace invariants are well-defined.
\end{remark}

Similarly to the continuous case the following proposition holds.
\begin{proposition}
Laplace transformation ${\mathcal L}\mapsto\hat{\mathcal L}$ defined by the ope\-ra\-tor ${\mathcal D}=\pa_x+b_{n-1}$ is invertible if
$a_n h_n\ne 0$. The inverse transformation $\hat{\mathcal L}\mapsto{\mathcal L}$ is given by the operator $-(a_n h_n)^{-1}(T_n+a_n)$. Similarly,
the transformation defined by the operator $T_n+a_n$ is invertible if $a_n k_n\ne 0$, and the inverse transformation in given by $-(a_n k_n)^{-1}(\pa_x+b_n)$.
\end{proposition}

\section{Darboux formulae in discrete case}\label{discr}

In this Section, we prove the analogs of the Darboux formulae for the general solution of linear hyperbolic equation with finite Laplace series in purely discrete case. Our aim is to prove the following
\begin{theorem}\label{thdarb}
Let ${\mathcal L}_0$ be a hyperbolic difference operator with finite Laplace series where $h_r=k_{-s}=0$ for some $r,s\geqslant 0$. Then there exist functions 
\begin{equation}
\label{func}
\nu_{1,n},\nu_{2,n},\dots,\nu_{r+s+1,n}\quad\hbox{and}\quad \mu_{1,m},\mu_{2,m},\dots,\mu_{r+s+1,m}
\end{equation}
of one discrete variable and a function $\omega_{nm}$ such that the general solution of ${\mathcal L}_0\psi=0$ is given by the determinant formula
\begin{multline}
\label{darb}
\psi_{0,n,m}=\\
=\omega_{n,m}\cdot\det\left(
\begin{array}{cccccccc}
M_m & M_{m+1} & \dots & M_{m+r} & N_n & N_{n+1} & \dots & N_{n+s}\\
\mu_{1,m} & \mu_{1,m+1} & \dots & \mu_{1,m+r} & \nu_{1,n} & \nu_{1,n+1} & \dots & \nu_{1,n+s}\\
\mu_{2,m} & \mu_{2,m+1} & \dots & \mu_{2,m+r} & \nu_{2,n} & \nu_{2,n+1} & \dots & \nu_{2,n+s}\\
\vdots &&&\vdots&&&&\vdots\\
\mu_{r+s+1,m} & \mu_{r+s+1,m+1} & \dots & \mu_{r+s+1,m+r} & \nu_{r+s+1,n} & \nu_{r+s+1,n+1} & \dots & \nu_{r+s+1,n+s}\\
\end{array}
\right),
\end{multline}
where $N_n$ and $M_m$ are arbitrary functions of one variable.\\

For any two families of linearly independent functions of one variable~(\ref{func}) and for any function $\omega_{n,m}$ there exists a hyperbolic difference operator 
${\mathcal L}_0$ with finite Laplace series that terminates at the terms $r$ and $-s$ such that formula~(\ref{darb}) provides the general solution of 
${\mathcal L}_0\psi=0$.
\end{theorem}

The proof of this general result is divided into a number of propositions, lemmas and theorems.
\begin{lemma}
If the Laplace series of ${\mathcal L}_0$ terminates at the term ${\mathcal L}_r$ where $r>0$, then the equation ${\mathcal L}_0\psi=0$ admits a solution of the form
\begin{equation}
\label{www}
\psi_{0,n,m}=B_{0,n,m} M_m+B_{1,n,m} M_{m+1}+\dots+B_{r,n,m} M_{m+r},
\end{equation}
where $M_m$ is an arbitrary function of one discrete variable. If the Laplace series terminates at the term ${\mathcal L}_{-s}$ where $s>0$, then the equation ${\mathcal L}_0\psi=0$ admits a solution of the form
\begin{equation}
\label{yyy}
\psi_{0,n,m}=A_{0,n,m} N_n+A_{1,n,m} N_{n+1}+\dots+A_{s,n,m} N_{n+s},
\end{equation}
where $M_m$ is an arbitrary function of one discrete variable.
\end{lemma}
{\bf Proof}.

If $h_r=0$, then the general solution of the equation ${\mathcal L}_r\psi=0$ has form~(\ref{zzz}). Remove the nonlocal part by setting $\psi_{n,0}=0$ for all $n\ne 0$, denote $M_m=\psi_{0,m}$ and use~(\ref{ttt}) to obtain solutions of ${\mathcal L}_0\psi=0$. Since a shift operator of order $r$ is applied to solutions of ${\mathcal L}_r\psi=0$, we get~(\ref{www}). Formula~(\ref{yyy}) is obtained similarly. $\Box$
\begin{remark}
\rm
Note that formulae~(\ref{www}) and~(\ref{yyy}) do not give the general solution of ${\mathcal L}_0\psi=0$.
\end{remark}
\begin{lemma}\label{le2}
If a difference hyperbolic equation ${\mathcal L}_0\psi=0$ admits a solution of form~(\ref{www}) where $M_m$ is an arbitrary function of one variable, then there exists a non-negative $p\leqslant r$ such that the Laplace series terminates: $h_p=0$. If ${\mathcal L}_0\psi=0$ admits a solution of form~(\ref{yyy}) where, then there exists a non-negative $p\leqslant s$ such that the Laplace series terminates: $k_{-p}=0$.
\end{lemma}
{\bf Proof}. Use induction in $r$ to prove the first claim. If $r=0$, then the equation ${\mathcal L}_0\psi=0$ admits a solution of the form 
$\psi_{n,m}=B_{0,n,m} M_m$. Substitute it into the equation:
$$
B_{0,n+1,m+1} M_{m+1}+a_{n,m} B_{0,n+1,m} M_m+b_{n,m} B_{0,n,m+1} M_{m+1}+c_{n,m}B_{0,n,m} M_m=0.
$$
Since $M$ is arbitrary, this leads to equations
\begin{eqnarray}
\nonumber
\left\{
\begin{array}{l}
B_{0,n+1,m+1}+b_{n,m} B_{0,n,m+1}=0\\
a_{n,m} B_{0,n+1,m}+c_{n,m}B_{0,n,m}=0
\end{array}
\right..
\end{eqnarray}
Multiply the first equation by $a_{n,m+1}$, shift the second equation in $m$ and consider the difference:
$$
a_{n,m+1}b_{n,m} B_{0,n,m+1}-c_{n,m+1}B_{0,n,m+1}=0
$$
This relation has to be satisfied for all namural $n,m$. Hence $a_{n,m+1}b_{n,m}-c_{n,m+1}=0$, which is equivalent to $h_{n,m+1}=0$.

Assume the proposition holds for all equations admitting solutions~(\ref{www}) of order $r$. Consider equation ${\mathcal L}_0\psi=0$ having such a solution of order $r+1$. This means that the result of substitution of this solution into the equation, which is an expression of the form
\begin{equation}
\label{abb}
C_{0,n,m}M_m+C_{1,n,m}M_{m+1}+\dots+C_{r+2,n,m}M_{m+r+2},
\end{equation}
identically vanishes. Note that the leading coefficient in~(\ref{abb}) has the form
\begin{equation}
\label{acc}
C_{r+2,n,m}=B_{r+1,n+1,m+1}+b_{n,m}B_{r+1,n,m+1}=0.
\end{equation}
If $h_0=0$, then the proposition is proved. Otherwise, apply the Laplace transformation $T_n+b_{n,m-1}$:
\begin{multline*}
\psi_{1,n,m}=(T_n+b_{n,m-1})\left(B_{0,n,m} M_m+B_{1,n,m} M_{m+1}+\dots+B_{r+1,n,m} M_{m+r+1}\right)=\\
=b_{n,m-1}B_{0,n,m} M_m+\dots+(B_{r+1,n+1,m}+b_{n,m-1}B_{r+1,n,m}) M_{m+r+1}
\end{multline*}
Compare the leading coefficient in the latter expression with~(\ref{acc}) to conclude that $\psi_1$ has form~(\ref{www}) of order $r$ and hence by the assumption the Laplace series of the corresponding operator ${\mathcal L}_1$ terminates at the term ${\mathcal L}_p$, where $p\leqslant r$. But 
${\mathcal L}_1$ is the next term in the Laplace series of the initial operator ${\mathcal L}_0$. Therefore its Laplace series terminates not later than at the 
$(r+1)$-th term. The second claim is proved similarly. $\Box$
\begin{lemma}\label{le3}
If the Laplace invariants of ${\mathcal L}_0$ are non-zero and the general solution of hyperbolic equation ${\mathcal L}_0\psi=0$ has the form
\begin{equation}
\label{vvv}
\psi_{0,n,m}=A_{0,n,m} N_n+A_{1,n,m} N_{n+1}+\dots+A_{s,n,m} N_{n+s}+B_{0,n,m} M_m+B_{1,n,m} M_{m+1}+\dots+B_{r,n,m} M_{m+r},
\end{equation}
where $M$ and $N$ are arbitrary functions of one discrete variable and $A_0\ne 0$, $A_s\ne 0$, $B_0\ne 0$, $B_r\ne 0$, then the general solution of the equation ${\mathcal L}_1\psi=0$, where 
${\mathcal L}_1=(T_n+b_{n,m-1}){\mathcal L}_0$, has the form
$$
\psi_{1,n,m}=\hat A_{0,n,m} N_n+\hat A_{1,n,m} N_{n+1}+\dots+\hat A_{s+1,n,m} N_{n+s+1}+\hat B_{0,n,m} M_m+\hat B_{1,n,m} M_{m+1}+\dots+
\hat B_{r-1,n,m} M_{m+r-1},
$$
where $\hat A_0\ne 0$, $\hat A_{s+1}\ne 0$, $\hat B_0\ne 0$, $\hat B_{r-1}\ne 0$, and the general solution of the equation ${\mathcal L}_{-1}\psi=0$, where ${\mathcal L}_{-1}=(T_m+a_{n-1,m}){\mathcal L}_0$, has the form
$$
\psi_{-1,n,m}=\tilde A_{0,n,m} N_n+\tilde A_{1,n,m} N_{n+1}+\dots+\tilde A_{s-1,n,m} N_{n+s-1}+\tilde B_{0,n,m} M_m+\tilde B_{1,n,m} M_{m+1}+
\dots+\tilde B_{r+1,n,m} M_{m+r+1},
$$
where $\tilde A_0\ne 0$, $\tilde A_{s-1}\ne 0$, $\tilde B_0\ne 0$, $\tilde B_{r+1}\ne 0$.
\end{lemma}
{\bf Proof}.

Apply the Laplace transformation to~(\ref{vvv}) to get the general solution of ${\mathcal L}_1\psi=0$. Obviously it has the form
$$
\psi_{1,n,m}=\hat A_{0,n,m} N_n+\hat A_{1,n,m} N_{n+1}+\dots+\hat A_{s+1,n,m} N_{n+s+1}+\hat B_{0,n,m} M_m+\hat B_{1,n,m} M_{m+1}+\dots+
\hat B_{r,n,m} M_{m+r},
$$
and we need to prove that $\hat B_{r,n,m}=0$. Note that
\begin{equation}
\label{add}
\hat A_{s+1,n,m}=A_{s,n+1,m},\quad\hat B_{r,n,m}=B_{r,n+1,m}+b_{n,m-1} B_{r,n,m},\quad\hat B_{r-1,n,m}=B_{r-1,n+1,m}+b_{n,m-1} B_{r-1,n,m};
\end{equation}
hence, $\hat A_{s+1}\ne 0$. Substitute~(\ref{vvv}) into hyperbolic equation ${\mathcal L}_0\psi=0$. This gives an expression of the form
$$
C_{0,n,m} N_n+C_{1,n,m} N_{n+1}+\dots+C_{s+1,n,m} N_{n+s+1}+ D_{0,n,m} M_m+D_{1,n,m} M_{m+1}+\dots+D_{r+1,n,m} M_{m+r+1}=0.
$$
Since functions $M$ and $N$ are arbitrary, all coefficients vanish. In particular,
$$
0=D_{r+1,n,m}=B_{r,n+1,m+1}+b_{n,m}B_{r,n,m+1}.
$$
Compare this with the second relation in~(\ref{add}) to verify that $\hat B_r=0$. Besides this,
\begin{equation}
\label{aee}
0=D_{r,n,m}=B_{r-1,n+1,m+1}+a_{n,m}B_{r,n+1,m}+b_{n,m}B_{r-1,n,m+1}+c_{n,m}B_{r,n,m}.
\end{equation}
Assume that $\hat B_{r-1}=0$. Hence~(\ref{add}) yields $B_{r-1,n+1,m+1}=-b_{n,m} B_{r-1,n,m+1}$. Therefore~(\ref{aee}) can be rewritten as 
$$
0=\left(c_{n,m}-a_{n,m}b_{n,m-1}\right)B_{r,n,m}=a_{n,m}b_{n,m-1}h_{n,m}B_{r,n,m},
$$
which contradicts the assumption $B_r\ne 0$ since $h$ is non-zero. 

Note that
$$
\hat A_{0,n,m}=b_{n,m-1}A_{0,n,m}\ne 0,\quad\hat B_{0,n,m}=B_{0,n+1,m}+b_{n,m-1}B_{0,n,m}.
$$
If $\hat B_{0,n,m}=0$, then express $B_{0,n+1,m}=-b_{n,m-1}B_{0,n,m}$ and substitute this into the equation $D_{0,n,m}=0$:
$$
0=a_{n,m}B_{0,n+1,m}+c_{n,m}B_{n,m}=\left(c_{n,m}-a_{n,m}b_{n,m-1}\right)B_{0,n,m}.
$$
Since $h\ne 0$, we arrive at a contradiction. The second claim is obtained similarly. $\Box$
\begin{theorem}\label{thsol}
If the Laplace series of ${\mathcal L}_0$ is finite, $h_r=k_{-s}=0$ for some $r,s\geqslant 0$ and the operators 
${\mathcal L}_{-s+1},\dots,{\mathcal L}_{r-1}$ have non-zero Laplace invariants, then the general solution of difference equation 
${\mathcal L}_0\psi=0$  is given by~(\ref{vvv}), where $N_n$ and $M_m$ are arbitrary functions of one variable, $A_{i,n,m}$ and $B_{j,n,m}$ are some particular functions depending the coefficients of the initial equation, and $A_0\ne 0$, $A_s\ne 0$, $B_0\ne 0$, $B_r\ne 0$.

If a hyperbolic equation ${\mathcal L}_0\psi=0$ admits a solution of the form~(\ref{vvv}), where $A_0\ne 0$, $A_s\ne 0$, $B_0\ne 0$ and $B_r\ne 0$, then its Laplace series is finite: $h_r=k_{-s}=0$ and the operators ${\mathcal L}_{-s+1},\dots,{\mathcal L}_{r-1}$ have non-zero Laplace invariants.
\end{theorem}
{\bf Proof}.

The general solution of ${\mathcal L}_0=0$ can be obtained by the application of $s$ Laplace transformations to the general solution of ${\mathcal L}_{-s}=0$. We will construct such a solution in a special form. Since $h_{-s}=0$, operator ${\mathcal L}_{-s}$ can be factorised and therefore according to Proposition~\ref{propsol} the general solution has the form $\psi_{-s,n,m}=\theta_{n,m}+\theta^0_{n,m}$. Note that $\theta_{n,m}=\alpha_{n,m}N_n$, where $N_n$ is arbitrary function of one variable and $\alpha_{n,m}$ depends on coefficients of ${\mathcal L}_{-s}$. Similarly, the general solution of the equation ${\mathcal L}_r=0$ has the form 
$\psi_{r,n,m}=\chi_{n,m}+\chi^0_{n,m}$,  where $\chi_{n,m}=\beta_{n,m}M_m$ and $M_m$ is an arbitrary function. Set $\chi^0=0$ and apply the composition ${\mathcal M}$ of inverse Laplace transformations:
$$
{\mathcal M}\chi_{n,m}=\omega_{n,m}\left(T_m+a_{-s+1,n-1,m}\left)\right(T_m+a_{-s+2,n-1,m}\right)\dots\left(T_m+a_{r,n-1,m}\right)\chi_{n,m}\in\Ker{\mathcal L}_{-s}
$$
(see Remark~\ref{reminv}). Since $\chi_{n,m}=\beta_{n,m}M_m$ depends on arbitrary function and 
$$
\left(T_n+b_{-s,n,m}\right)\left(T_m+a_{-s,n-1,m}\right){\mathcal M}\chi_{n,m}=0,
$$
the function $\phi_{n,m}=\left(T_m+a_{-s,n-1,m}\right){\mathcal M}\chi_{n,m}$ parameterises the general solution of homogeneous equation 
$\left(T_n+b_{-s,n,m}\right)\psi_{n,m}=0$. Hence, $\theta^0_{n,m}={\mathcal M}\chi_{n,m}$ satisfies non-homogeneous equation 
$$
\left(T_m+a_{-s,n-1,m}\right)\psi_{n,m}=\phi_{n,m}.
$$
Therefore the general solution of ${\mathcal L}_{-s}\psi=0$ has the form
$$
\psi_{-s,n,m}=\alpha_{n,m}N_n+{\mathcal M}\chi_{n,m}.
$$
Apply the Laplace transformations ${\mathcal N}=\left(T_n+b_{-1,n,m-1}\right)\dots\left(T_n+b_{-s,n,m-1}\right)$ to obtain the general solution of 
${\mathcal L}_0\psi=0$:
\begin{multline*}
\psi_{0,n,m}={\mathcal N}\left(\alpha_{n,m}N_n+{\mathcal M}\chi_{n,m}\right)=
{\mathcal N}\left(\alpha_{n,m}N_n\right)+\tilde\omega_{n,m}\left(T_m+a_{1,n-1,m}\right)\dots\left(T_m+a_{r,n-1,m}\right)\left(\beta_{n,m}M_m\right)=\\
=A_{0,n,m} N_n+A_{1,n,m} N_{n+1}+\dots+A_{s,n,m} N_{n+s}+B_{0,n,m} M_m+B_{1,n,m} M_{m+1}+\dots+B_{r,n,m} M_{m+r}
\end{multline*}
since ${\mathcal N}$ and $\left(T_m+a_{1,n-1,m}\right)\dots\left(T_m+a_{r,n-1,m}\right)$ are difference operators in $T_n$ and $T_m$ of orders $s$ and $r$ respectively.

The second claim follows from Lemmas~\ref{le2},~\ref{le3}. $\Box$
\smallskip

Suppose the Laplace series of ${\mathcal L}_0$ is finite, $h_r=k_{-s}=0$ for some $r,s\geqslant 0$ and the operators 
${\mathcal L}_{-s+1},\dots,{\mathcal L}_{r-1}$ have non-zero Laplace invariants. Then the general solution of the equation ${\mathcal L}_r\psi=0$ has the form
$$
\psi_{r,n,m}=A_{0,n,m} N_n+A_{1,n,m} N_{n+1}+\dots+A_{r+s,n,m} N_{n+r+s}+B_{0,n,m} M_m.
$$
Without loss of generality one may assume that $B_0=1$ (apply an appropriate gauge transformation, which does not change the Laplace invariants). Hence, the hyperbolic equation ${\mathcal L}_r\psi=0$ admits a solution of the form $\psi_{r,n,m}=M_m$. This yields $b_{r,n,m}=-1$, $a_{r,n,m}=-c_{r,n,m}$. 
Thus, 
$$
{\mathcal L}_r=\left(T_m+a_{r,n,m}\right)\left(T_n-1\right)\quad\hbox{and}\quad\left(T_n-1\right)\psi_{r,n,m}\in\Ker\left(T_m+a_{r,n,m}\right).
$$ 
Therefore, $\left(T_n-1\right)\psi_{r,n,m}=\alpha_{n,m}\tilde N_n$ for a certain $\tilde N$. This leads to the relation
\begin{multline}
\label{agg}
-A_{0,n,m} N_n+(A_{0,n+1,m}-A_{1,n,m}) N_{n+1}+\dots+\\
+(A_{r+s-1,n+1,m}-A_{r+s,n,m}) N_{n+r+s}+A_{r+s,n+1,m} N_{n+r+s+1}=\alpha_{n,m}\tilde N_n.
\end{multline}
Denote
$$
\lambda_{0,n,m}=-\frac{A_{0,n,m}}{\alpha_{n,m}},\quad\lambda_{1,n,m}=\frac{A_{0,n+1,m}-A_{1,n,m}}{\alpha_{n,m}},\dots,
\lambda_{r+s+1,n,m}=\frac{A_{r+s,n+1,m}}{\alpha_{n,m}}.
$$
Note that all coefficients $\lambda_0,\dots,\lambda_{r+s+1}$ do not depend on $m$ since $N_n$ and $\tilde N_n$ are functions of one discrete variable $n$. Therefore we drop the second index in $\lambda_{0,n,m},\dots,\lambda_{r+s+1,n,m}$ for convenience.
\begin{lemma}
Function $\alpha_{n,m}$ satisfies a linear difference equation of order $r+s+1$ in the variable $n$ with coefficients depending only on $n$.
\end{lemma}
{\bf Proof}.

Express the functions $A_{i,n,m}$ in terms of $\alpha_{n,m}$ and $\lambda_0,\dots,\lambda_{r+s+1}$:
\begin{align*}
A_{0,n,m}&=-\alpha_{n,m}\lambda_{0,n}\\
A_{1,n,m}&=A_{0,n+1,m}-\alpha_{n,m}\lambda_{1,n}=-T_n\left(\alpha_{n,m}\lambda_{0,n}\right)-\alpha_{n,m}\lambda_{1,n}\\
A_{2,n,m}&=A_{1,n+1,m}-\alpha_{n,m}\lambda_{2,n}=-T^2_n\left(\alpha_{n,m}\lambda_{0,n}\right)-T_n\left(\alpha_{n,m}\lambda_{1,n}\right)-\alpha_{n,m}\lambda_{2,n}\\
\dots\\
A_{r+s,n,m}&=A_{r+s-1,n+1,m}-\alpha_{n,m}\lambda_{r+s,n}=-\sum\limits_{i=0}^{r+s}T^i_n\left(\alpha_{n,m}\lambda_{r+s-i,n}\right)\\
A_{r+s,n+1,m}&=\alpha_{n,m}\lambda_{r+s+1,n}.
\end{align*}
Hence, function $\alpha_{n,m}$ satisfies linear difference equation
\begin{equation}
\label{diffeq}
\lambda_{0,n} T^{r+s+1}_n\alpha_{n,m}+\dots+\lambda_{r+s} T_n\alpha_{n,m}+\lambda_{r+s+1}\alpha_{n,m}=0.\quad\Box
\end{equation}

Let $\nu_{1,n},\nu_{2,n},\dots,\nu_{r+s+1,n}$ be a fundamental system of solutions of linear ordinary difference equation~(\ref{diffeq}). Then $\tilde N=0$ once $N=\nu_i$ is chosen in~(\ref{agg}) for some $i=1,2,\dots,r+s+1$. Hence
$$
\left(T_n-1\right)\left(A_{0,n,m} \nu_{i,n}+A_{1,n,m} \nu_{i,n+1}+\dots+A_{r+s,n,m} \nu_{i,n+r+s}+M_m\right)=0
$$
for every $i=1,2,\dots,r+s+1$. This means that
$$
\mu_{i,m}:=A_{0,n,m} \nu_{i,n}+A_{1,n,m} \nu_{i,n+1}+\dots+A_{r+s,n,m} \nu_{i,n+r+s},\quad i=1,2,\dots,r+s+1,
$$
are functions only of $m$.
\begin{proposition}
In the above setting the gerenal solution of the equation ${\mathcal L}_r\psi=0$ is given by $\psi_r=\frac{\Delta_1}{\Delta}$, where
\begin{eqnarray}
\label{ahh}
\Delta_{1,n,m}=\det\left(
\begin{array}{ccccc}
M_m & N_n & N_{n+1} & \dots &  N_{n+r+s}\\
\mu_{1,m} & \nu_{1,n} & \nu_{1,n+1} & \dots & \nu_{1,n+r+s}\\
\mu_{2,m} & \nu_{2,n} & \nu_{2,n+1} & \dots & \nu_{2,n+r+s}\\
\vdots &&&\ddots&\vdots\\
\mu_{r+s+1,m} & \nu_{r+s+1,n} & \nu_{r+s+1,n+1} & \dots & \nu_{n+s+1,n+r+s}
\end{array}
\right),
\end{eqnarray}
\begin{eqnarray}
\label{ajj}
\Delta_{n,m}=\det\left(
\begin{array}{cccc}
\nu_{1,n} & \nu_{1,n+1} & \dots & \nu_{1,n+r+s}\\
\nu_{2,n} & \nu_{2,n+1} & \dots & \nu_{2,n+r+s}\\
\vdots &&\ddots&\vdots\\
\nu_{r+s+1,n} & \nu_{r+s+1,n+1} & \dots & \nu_{n+s+1,n+r+s}
\end{array}
\right).
\end{eqnarray}
\end{proposition}
{\bf Proof}. Let $\psi_r$ be the general solution of ${\mathcal L}_r\psi=0$. Then there exist families of functions $\nu_1,\dots,\nu_{r+s+1}$ and 
$\mu_1,\dots,\mu_{r+s+1}$ depending on $n$ and $m$ respectively such that the following relations are satisfied:
\begin{equation}
\label{akk}
\left\{
\begin{aligned}
\psi_{r,n,m}&-A_{0,n,m} N_n-A_{1,n,m} N_{n+1}-\dots-A_{r+s,n,m} N_{n+r+s}=M_m\\
&-A_{0,n,m}\nu_{i,n}-A_{1,n,m} \nu_{i,n+1}-\dots-A_{r+s,n,m} \nu_{i,n+r+s}=\mu_{i,m},\quad i=1,2,\dots,r+s+1\\
\end{aligned}
\right..
\end{equation}
Solve this system with respect to the unknowns $\psi_r,-A_0,-A_1,\dots,-A_{r+s}$ and apply Cramer's rule to obtain the formula for $\psi_r$. $\Box$

Consider the general solution of ${\mathcal L}_0\psi=0$. On the one hand it has form~(\ref{vvv}); on the other hand it is obtained from the general solution of equation ${\mathcal L}_r\psi=0$ by application of Laplace transformations: $\psi_r={\mathcal M}\psi_0$, where ${\mathcal M}$ is a difference operator of order $r$  and $\psi_r$ is the quotient of determinants~(\ref{ahh}) and~(\ref{ajj}). Hence $\psi_0=0$ whenever $N=\nu_i$ and $M=\mu_i$ for a certain 
$i=1,2,\dots,r+s+1$. Therefore, up to multiplication by an appropriate function $\omega_{n,m}$, the following equations are satisfied:
$$
\left\{
\begin{aligned}
\psi_{0,n,m}&-B_{0,n,m} M_n-\dots-B_{r,n,m} M_{n+r}-A_{0,n,m} N_n-\dots-A_{s,n,m} N_{n+s}=0\\
&-B_{0,n,m}\mu_{i,n}-\dots-B_{r,n,m} \mu_{i,n+r}-A_{0,n,m}\nu_{i,n}-\dots-A_{s,n,m} \nu_{i,n+s}=0,\quad i=1,2,\dots,r+s+1\\
\end{aligned}
\right..
$$
Solve this system with respect to the unknowns $\psi_0,-B_0,\dots,-B_r,-A_0,\dots,A_s$ and apply the projective version of Cramer's rule to 
obtain~(\ref{darb}). This proves the first part of Theorem~\ref{thdarb}.

In order to prove the second part of Theorem~\ref{thdarb} choose arbitrary independent functions of one variable
\begin{equation}
\label{amm}
\nu_{1,n},\nu_{2,n},\dots,\nu_{r+s+1,n}\quad\hbox{and}\quad \mu_{1,m},\mu_{2,m},\dots,\mu_{r+s+1,m}
\end{equation}
and solve linear system~(\ref{akk}) for $A_0,A_1,\dots,A_{r+s}$. This provides 
\begin{equation}
\label{all}
\psi_{r,n,m}=A_{0,n,m} N_n+A_{1,n,m} N_{n+1}+\dots+A_{r+s,n,m} N_{n+r+s}+M_m.
\end{equation}
\begin{lemma}
Function~(\ref{all}) satisfies a second order hyperbolic equation ${\mathcal L}_r\psi=0$ such that $h_r=0$ for the operator ${\mathcal L}_r$.
\end{lemma}
{\bf Proof}. Since $\mu_{i,m}=-A_{0,n,m}\nu_{i,n}-A_{1,n,m} \nu_{i,n+1}-\dots-A_{r+s,n,m} \nu_{i,n+r+s}$ is a function of $m$ for every 
$\quad i=1,2,\dots,r+s+1$, function $\nu_i$ satisfies the difference equation
$$
\left(T_n-1\right)\left(A_{0,n,m} \nu_{i,n}+A_{1,n,m} \nu_{i,n+1}+\dots+A_{r+s,n,m} \nu_{i,n+r+s}\right)=0,
$$
which has order $r+s+1$. Consider linear ordinary difference equation
$$
\lambda_{0,n}\nu_{n+r+s+1}+\dots+\lambda_{r+s,n}\nu_{n+1}+\lambda_{r+s+1,n}\nu_n=0,
$$
having $\nu_{1,n},\nu_{2,n},\dots,\nu_{r+s+1,n}$ as a basis in the space of solutions. Such equation is unique up to multiplication by an arbitrary function 
$\alpha_{n,m}$. Therefore the following relations are satisfied:
$$
\alpha_{n,m}=-\frac{A_{0,n,m}}{\lambda_{0,n,m}}=\frac{A_{0,n+1,m}-A_{1,n,m}}{\lambda_{1,n}}=\dots=
\frac{A_{r+s-1,n+1,m}-A_{r+s,n,m}}{\lambda_{r+s,n}}=\frac{A_{r+s,n+1,m}}{\lambda_{r+s+1,n}}.
$$
This implies that $\left(T_n-1\right)\psi_{r,n,m}=\alpha_{n,m}\tilde N_n$, where $\tilde N$ is a function of $n$. Hence
$$
\left(T_m+\gamma_{n,m}\right)\left(T_n-1\right)\psi_{r,n,m}=\left(T_m+\gamma_{n,m}\right)\left(\alpha_{n,m}\tilde N_n\right)=
\left(\alpha_{n,m+1}+\gamma_{n,m}\alpha_{n,m}\right)\tilde N_n
$$
for any $\gamma_{n,m}$. Therefore if $\gamma_{n,m}=-\frac{\alpha_{n,m+1}}{\alpha_{n,m}}$ for all $n,m\in\mathbb Z$, then ${\mathcal L}_r\psi_r=0$, where
$$
{\mathcal L}_r=\left(T_m+\gamma_{n,m}\right)\left(T_n-1\right)=T_nT_m+\gamma_{n,m}T_n-T_m-\gamma_{n,m}.
$$
Note that $h_r=0$ for any operator of such kind. $\Box$
\smallskip

According to Theorem~\ref{thsol} if hyperbolic equation ${\mathcal L}_r\psi=0$ admits a solution of form~(\ref{all}), then the Laplace series of 
${\mathcal L}_r$ is finite. In particular, $h_{-s}=0$. Apply Laplace transformations to ${\mathcal L}_r$ and to $\psi_r$ in order to obtain the operator 
${\mathcal L}_0$ in the Laplace series and the general solution of the equation 
${\mathcal L}_0\psi=0$. On the one hand it has the form
$$
\psi_{0,n,m}=C_{0,n,m} N_n+C_{1,n,m} N_{n+1}+\dots+C_{s,n,m} N_{n+s}+D_{0,n,m} M_m+D_{1,n,m} M_{m+1}+\dots+D_{r,n,m} M_{m+r}.
$$
On the other hand consider the determinant in~(\ref{darb}) constructed using functions~(\ref{amm}) and expand it along its first row. This expression has the form
$$
\tilde\psi_{0,n,m}=\tilde A_{0,n,m} N_n+\tilde A_{1,n,m} N_{n+1}+\dots+\tilde A_{s,n,m} N_{n+s}+\tilde B_{0,n,m} M_m+\tilde B_{1,n,m} M_{m+1}+\dots+\tilde B_{r,n,m} M_{m+r}.
$$
Note that since $\psi_0$ is obtained from $\psi_r$, which was constructed earlier, the substitution $N=\nu_i$, $M=\mu_i$ impies that $\psi_r=0$ for every 
$i=1,2,\dots,r+s+1$; therefore we get $\psi_0=0$ for such a choice of $N$ and $M$ as well. This leads to linear system
$$
C_{0,n,m} \nu_{i,n}+C_{1,n,m} \nu_{i,n+1}+\dots+C_{s,n,m} \nu_{i,n+s}+D_{0,n,m} \mu_{i,m}+D_{1,n,m} \mu_{i,m+1}+\dots+D_{r,n,m} \mu_{i,m+r}=0
$$
for the unknowns $C_1,\dots,C_{s,n,m},D_1,\dots,D_{r,n,m}$. Note also that such choice of $N$ and $M$ yields $\tilde\psi_0=0$ and, therefore, also the relations
$$
\tilde A_{0,n,m} \nu_{i,n}+\tilde A_{1,n,m} \nu_{i,n+1}+\dots+\tilde A_{s,n,m} \nu_{i,n+s}+
\tilde B_{0,n,m} \mu_{i,m}+\tilde B_{1,n,m} \mu_{i,m+1}+\dots+\tilde B_{r,n,m} \mu_{i,m+r}=0,
$$
where $i=1,2,\dots,r+s+1$. Hence, the unknowns
$$
C_1,\dots,C_{s,n,m},D_1,\dots,D_{r,n,m}\quad\hbox{and}\quad\tilde A_1,\dots,\tilde A_{s,n,m},\tilde B_1,\dots,\tilde B_{r,n,m}
$$
satisfy the same system of linear homogeneous equations, and these $(r+s+1)$-tuples are proportional since functions~(\ref{amm}) are independent. Therefore there exists $\tilde\omega$ such that $\psi_{0,n,m}=\tilde\omega_{n,m}\tilde\psi_{0,n,m}$ defines the general solution of ${\mathcal L}_0\psi=0$, where 
${\mathcal L}_0$ has finite Laplace series. For any other $\omega$ function $\omega_{n,m}\psi_{0,n,m}$ satisfies equation 
$\tilde{\mathcal L}_0\psi=0$, where $\tilde{\mathcal L}_0=\frac{1}{\omega_{n+1,m+1}}{\mathcal L}_0\omega_{n,m}$ is gauge-equivalent to 
${\mathcal L}_0$ and therefore has the same series of Laplace invariants. This completes the proof of Theorem~\ref{thdarb}. $\Box$

In the discrete case it is also possible to express in terms of functions~(\ref{func}) not only the general solution~(\ref{darb}), but also the coefficients of the operator ${\mathcal L}_0$. The functions $A_{0,n,m}$ and $B_{0,n,m}$ can be expressed in terms of~(\ref{func}) using the expansion of 
determinant~(\ref{darb}) along its first row, see Remark~\ref{remcoef}. The coefficients of ${\mathcal L}_0$ can be expressed in terms of them.
\begin{proposition}
If the general solution of the equation $L_0\psi=0$ with the Laplace series terminating at the terms $r$ and $-s$ is given by~(\ref{darb}), 
then the coefficients of the operator ${\mathcal L}_0$ have the form
\begin{equation}
\label{app}
a_{n,m}=-\frac{A_{s,n+1,m+1}}{A_{s,n+1,m}},\ b_{n,m}=-\frac{B_{r,n+1,m+1}}{B_{r,n,m+1}},\ 
c_{n,m}=\frac{\sum\limits_{i=0}^s A_{i,n+1,m+1}+a_{nm}\sum\limits_{i=0}^s A_{i,n+1,m}+b_{nm}\sum\limits_{i=0}^s A_{i,n,m+1}}{\sum\limits_{i=0}^s A_{i,n,m}}.
\end{equation}
\end{proposition}
{\bf Proof}.

Consider the general solution of ${\mathcal L}_0\psi=0$ in form~(\ref{vvv}) and apply shifts:
\begin{align*}
\psi_{0,n+1,m}&=A_{0,n+1,m} N_{n+1}+\dots+A_{s,n+1,m} N_{n+s+1}+B_{0,n+1,m} M_m+\dots+B_{r,n+1,m} M_{m+r}\\
\psi_{0,n,m+1}&=A_{0,n,m+1} N_{n}+\dots+A_{s,n,m+1} N_{n+s}+B_{0,n,m+1} M_{m+1}+\dots+B_{r,n,m+1} M_{m+r+1}\\
\psi_{0,n+1,m+1}&=A_{0,n+1,m+1} N_{n+1}+\dots+A_{s,n+1,m+1} N_{n+s+1}+B_{0,n+1,m+1} M_{m+1}+\dots+B_{r,n+1,m+1} M_{m+r+1}.\\
\end{align*}
Note that the expression
$$
\tilde\psi_{0,n,m}:=\psi_{0,n+1,m+1}-\frac{A_{s,n+1,m+1}}{A_{s,n+1,m}}\psi_{0,n+1,m}-\frac{B_{r,n+1,m+1}}{B_{r,n,m+1}}\psi_{0,n,m+1}
$$
does not contain the highest shifts $N_{n+s+1}$ and $M_{m+r+1}$, i.e. it has form~(\ref{vvv}). Moreover, the substitution $N=\nu_i$, $M=\mu_i$ for every 
$i=1,2,\dots,r+s+1$ implies that $\psi_0=0$ and, hence, that $\tilde\psi_0=0$. Using the same argument as before we deduce that there exists a function 
$\omega$ such that  $\tilde\psi_{0,n,m}=\omega_{n,m}\psi_{0,n,m}$. Hence,
\begin{equation}
\label{ann}
\psi_{0,n+1,m+1}-\frac{A_{s,n+1,m+1}}{A_{s,n+1,m}}\psi_{0,n+1,m}-\frac{B_{r,n+1,m+1}}{B_{r,n,m+1}}\psi_{0,n,m+1}=\omega_{n,m}\psi_{0,n,m}.
\end{equation}
This means that we constructed a hyperbolic second order difference operator with unitary leading coefficient that is annihilated by the general solution $\psi_0$ of the initial equation. Hence we only need to prove the formula for $c_{n,m}=-\omega_{n,m}$. Consider particular solution that corresponds to $N_n\equiv 1$, $M_m\equiv 0$ and substitute it into~(\ref{ann}) to find $\omega$. $\Box$
\begin{remark}
\rm
It is also possible to obtain a formula for $c_{n,m}$ in~(\ref{app}) in terms of coefficients $B_{j,n,m}$. This corresponds to another choice of a particular solution. Moreover, some other particular solution would give yet another formula for $c_{n,m}$. There is no contradiction in such inambiguity since the coefficients $A_{i,n,m}$, $B_{j,n,m}$ satisfy strong restrictions imposed by formula~(\ref{darb}).
\end{remark}

\section{Semi-discrete case}\label{semidiscr}

In this Section, we briefly discuss semi-discrete analogs of the Darboux formulae. All proofs are similar to the ones in the continuous and in the entirely discrete cases.
\begin{theorem}
If the Laplace series of a semi-discrete hyperbolic operator ${\mathcal L}_0$ is finite, $h_r=k_{-s}=0$ for some $r,s\geqslant 0$ and the operators 
${\mathcal L}_{-s+1},\dots,{\mathcal L}_{r-1}$ have non-zero Laplace invariants, then the general solution of difference equation 
${\mathcal L}_0\psi=0$ has the form
$$
\psi_{0,n}=A_{0,n} N_n+A_{1,n} N_{n+1}+\dots+A_{r,n} N_{n+r}+B_{0,n} X+B_{1,n} X'+\dots+B_{s,n} X^{(s)},
$$
where $N_n$ and $X$ are arbitrary functions of variables $n$ and $x$ respectively, $A_{i,n}$ and $B_{j,n}$ are some particular functions depending the coefficients of the initial equation, and $A_0\ne 0$, $A_r\ne 0$, $B_0\ne 0$, $B_s\ne 0$.

If a hyperbolic equation ${\mathcal L}_0\psi=0$ admits a solution of form~(\ref{vvv}), where $A_0\ne 0$, $A_r\ne 0$, $B_0\ne 0$ and $B_s\ne 0$, then its Laplace series is finite, $h_r=k_{-s}=0$, and the operators ${\mathcal L}_{-s+1},\dots,{\mathcal L}_{r-1}$ have non-zero Laplace invariants.
\end{theorem}
\begin{theorem}
Let ${\mathcal L}_0$ be a hyperbolic differential-difference operator with finite Laplace series where $h_r=k_{-s}=0$ for some $r,s\geqslant 0$. Then there exist functions 
\begin{equation}
\label{sdfunc}
\nu_{1,n},\nu_{2,n},\dots,\nu_{r+s+1,n}\quad\hbox{and}\quad \xi_1 (x),\xi_2 (x),\dots,\xi_{r+s+1} (x)
\end{equation}
of one discrete variable and a function $\omega_n=\omega_n(x)$ such that the general solution of ${\mathcal L}_0\psi=0$ is given by the determinant formula
\begin{eqnarray}
\label{sddarb}
\psi_{0,n}=\omega_{n}\cdot\det\left(
\begin{array}{cccccccc}
X & X' & \dots & X^{(s)} & N_n & N_{n+1} & \dots & N_{n+r}\\
\xi_1 & \xi'_1 & \dots & \xi^{(s)}_1 & \nu_{1,n} & \nu_{1,n+1} & \dots & \nu_{1,n+r}\\
\xi_2 & \xi'_2 & \dots & \xi^{(s)}_2 & \nu_{2,n} & \nu_{2,n+1} & \dots & \nu_{2,n+r}\\
\vdots &&&\vdots&&&&\vdots\\
\xi_{r+s+1} & \xi'_{r+s+1} & \dots & \xi^{(s)}_{r+s+1} & \nu_{r+s+1,n} & \nu_{r+s+1,n+1} & \dots & \nu_{r+s+1,n+r}\\
\end{array}
\right),
\end{eqnarray}
where $N_n$ and $X(x)$ are arbitrary functions of one variable.\\

For any two families of linearly independent functions of one variable~(\ref{sdfunc}) and for any function $\omega_n$ there exists a hyperbolic differential-difference operator ${\mathcal L}_0$ with finite Laplace series that terminates at the terms $r$ and $-s$ such that formula~(\ref{sddarb}) provides the general solution of ${\mathcal L}_0\psi=0$.
\end{theorem}

\section{Acknowledgements}

The author is grateful to Prof.~Alexander Bobenko for stimulating discussions and for hospitality during his stay in TU Berlin where the work had been started. The work on the paper was completed during the author's stay in the University of Glasgow within the frame of INI Solidarity Satelline Programme for Mathematicians. The author wishes to thank the Isaac Newton Institute and the Programme for financial support during his stay in Glasgow.

\end{document}